\documentclass[
 reprint,
 amsmath,amssymb,
prl,
]{revtex4-2}

\usepackage{color}
\usepackage{rotate}
\usepackage{graphicx}
\usepackage{dcolumn}
\usepackage{bm}

\newcommand\Nu{\text{Nu}}
\newcommand\Ra{\text{Ra}}

\newcommand\Pran{\text{Pr}}

\newcommand\Ek{\text{Ek}}
\newcommand\Ro{\text{Ro}}

\begin{document}

\title{Connecting wall modes and boundary zonal flows in rotating Rayleigh--B\'enard convection}

\author{Robert~E.~Ecke$^{1,2,3}$}
\email{ecke@lanl.gov}
\author{Xuan~Zhang$^{1}$}
\email{xuan.zhang@ds.mpg.de}
\author{Olga~Shishkina$^{1}$}
\email{Olga.Shishkina@ds.mpg.de}
\affiliation{$^{1}$Max Planck Institute for Dynamics and Self-Organization, 37077 G\"ottingen, Germany }
\affiliation{$^{2}$Center for Nonlinear Studies, Los Alamos National Laboratory, Los Alamos, New Mexico 87545, USA}
\affiliation{$^{3}$Department of Physics, University of Washington, Seattle, WA 98195, USA}

\date{\today}

\begin{abstract}
Using direct numerical simulations, we study rotating Rayleigh-B\'enard convection in a cylindrical cell for a broad range of Rayleigh, Ekman, and Prandtl numbers from the onset of wall modes to the geostrophic regime, an extremely important one in geophysical and astrophysical contexts. We connect linear wall-mode states that occur prior to the onset of bulk convection with the boundary zonal flow that coexists with turbulent bulk convection in the geostrophic regime through the continuity of length and time scales and of convective heat transport. We quantitatively collapse drift frequency, boundary length, and heat transport data from numerous sources over many orders of magnitude in Rayleigh and Ekman numbers. Elucidating the heat transport contributions of wall modes and of the boundary zonal flow are critical for characterizing the properties of the geostrophic regime of rotating convection in finite, physical containers and is crucial for connecting the geostrophic regime of laboratory convection with geophysical and astrophysical systems.
\end{abstract}

\maketitle

Rayleigh--B\'enard convection with rotation (RRBC) about a vertical axis is a prototypical laboratory realization of geophysical and astrophysical systems that combines buoyancy forcing and rotation \cite{Chandrasekhar1961,Rossby1969,Pfotenhauer1984,Zhong1993,Ning1993,Julien1996,Liu1999,King2009,Zhong2009}.  Much recent experimental \cite{King2009,Ecke2014,Kunnen2021,Wedi2021}  and theoretical/numerical interest \cite{Sprague2006,Julien2012,Stellmach2014} in rotating convection has focused on the geostrophic regime where  rotation dominates.  In particular, one is interested in the scaling of the normalized heat transport $\Nu$ with $\Ra$ to compare with theoretical predictions of asymptotic models that provide insight into broader geo- and astrophysical situations.  There are significant experimental challenges \cite{Cheng2018} for making a compelling comparison including reaching small $\Ek$ number with correspondingly large $\Ra$.  Consequently, the geometry of experimental convection cells have tended towards small aspect ratio $\Gamma = D/H < 1$, where $D$ and $H$ are the cell diameter and height, respectively.  Recently several investigations \cite{Zhang2020,Zhang2021,Wit2020} have revealed a boundary zonal flow (BZF) that contributes strongly to total heat transport. The BZF has features reminiscent of wall mode states in RRBC \cite{Zhong1993,Herrmann1993,Kuo1993,Liu1999} and a numerical study \cite{Favier2020} indicated that the BZF was indeed the nonlinear remnant of wall modes.  Here we establish unambiguously through direct numerical simulation (DNS) and comparison among disparate data sets for the drift frequency $\omega_d$, the radial length scale $\delta_0$, and the heat transport $\Nu$, that there is a continuous evolution of the wall mode states into the BZF which coexists with the geostrophic convection modes. We also find that the forgotten wall mode contribution to the heat transport plays an important role in determining the scaling of $\Nu$ in the geostrophic regime, a crucial element in a proper comparison among experiment, DNS, and theory.   

\begin{figure}\centering
\includegraphics[width=8.5cm]{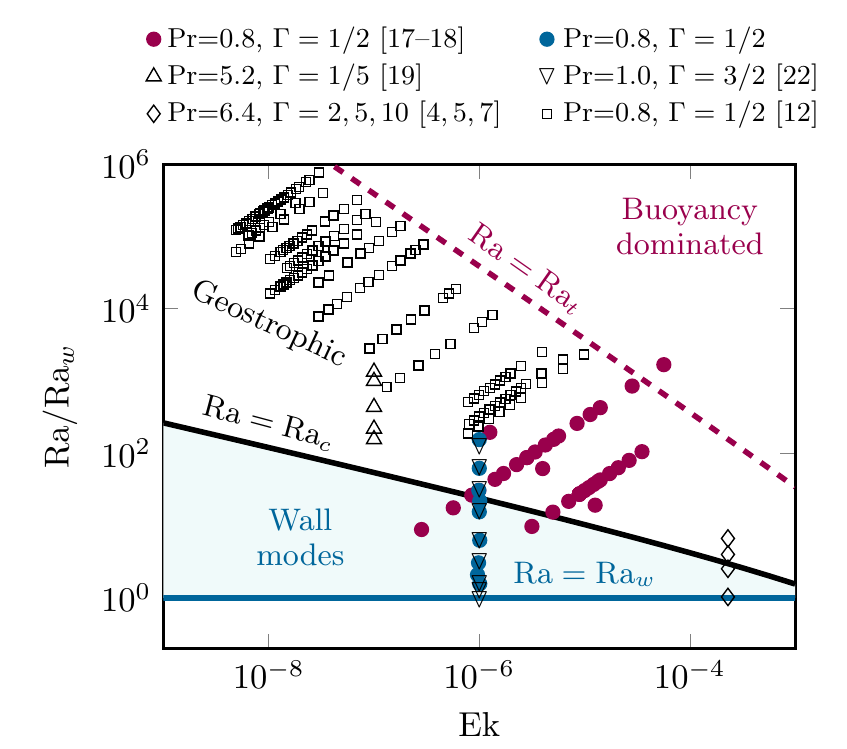}
\caption{Phase diagram of states of rotating Rayleigh--B\'enard convection: $\Ra/\Ra_w$ vs. $\Ek$. 
Symbols for different data sets: new data reported here (solid circle - red), 
\cite{Zhang2020, Zhang2021} (solid circle - blue), 
\cite{Wit2020} (solid square - black), 
\cite{Favier2020} (solid triangles - black), 
\cite{Wedi2021} (open squares - black), 
and \cite{Zhong1993} (open diamonds - black). 
}
\label{PIC1}
\end{figure}

The dimensionless control parameters in RRBC are the Rayleigh number $\Ra = \alpha g \Delta H^3/(\kappa \nu)$, Prandtl number $\Pran = \nu/\kappa$, Ekman number $\Ek = \nu/(2\Omega H^2)$, and cell aspect ratio $\Gamma$ where $\alpha$ is isobaric thermal expansion coefficient, $\nu$ kinematic viscosity, $\kappa$ fluid thermal diffusivity, $g$  acceleration of gravity, $\Omega$ angular rotation rate, and $\Delta$ the temperature difference between horizontal confining plates. The global response of the system is the normalized heat transport $\Nu$, and the time and length scales of the wall modes and of the BZF are the normalized precession frequency $\omega_d = \omega/\Omega$ and radial localization length scale $\delta_0/H$.  We present data for $\Ek = 10^{-6}$, $\Pran=0.8$, $\Gamma = 1/2$, and $2 \times 10^7 \leq \Ra \leq 5 \times 10^9$ that spans the wall mode onset at $Ra_w = 2.8 \times 10^7$ through the onset of bulk convection at $\Ra_c \approx 9 \times 10^8$.  We use our results on this system over wider ranges of $\Ek$ and $\Ra$ \cite{Zhang2020,Zhang2021} with data from other experiments and DNS \cite{Zhong1993,Ning1993,Liu1999,Wit2020,Favier2020,Wedi2021} to test our proposed power-law scalings.  
 
The regimes of rotating convection in finite containers are wall-mode states at the lowest $\Ra$, followed by a transition to the geostrophic state of rotating convection, and finally to a transition to weakly-rotating states at the highest $\Ra$. The first instability from the no-convection base state is to wall modes with critical Rayleigh number $\Ra_w \approx 31.8 \Ek^{-1} + 46.6 \Ek^{-2/3}$ \cite{Herrmann1993,Zhang2009}.  
\begin{figure}[h]
\includegraphics[width=8.5cm]{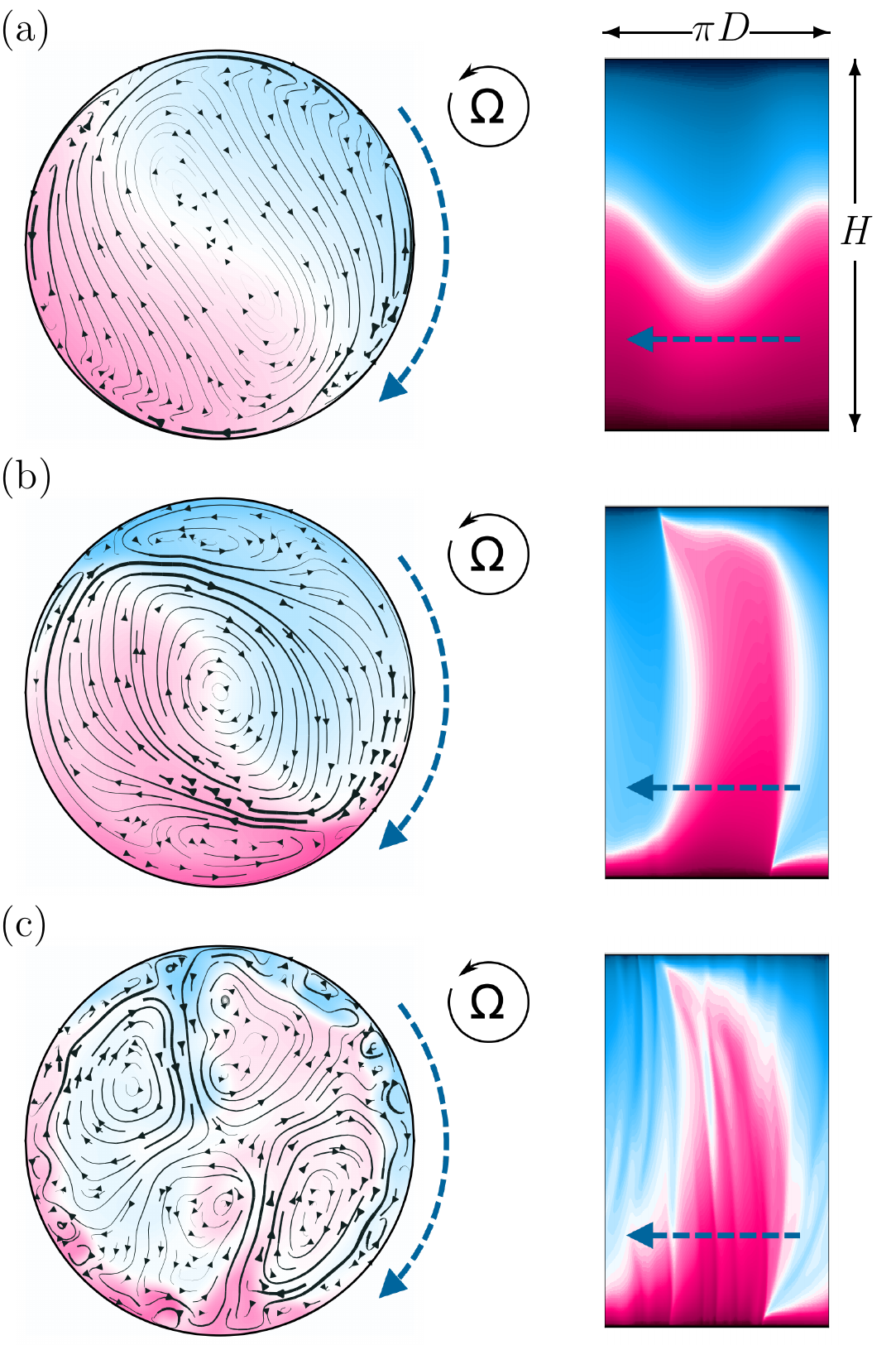}
\caption{Instantaneous temperature fields (left -- horizontal at $z=H/2$ with streamlines; right -- vertical at $r=0.98R$) for $\Ek = 10^{-6}$. 
Corresponding $\Ra$ and $\epsilon$ and $\Ro$:
(a) $3 \times 10^7$, 0.071, (b) $5 \times 10^8$, 17, (c) $1 \times 10^9$, 35. Directions of rotation and wall mode precession are shown. 
}
\label{PIC2}
\end{figure}
To emphasize the role of these wall modes, we plot the boundaries of rotating convection regimes in Fig.~\ref{PIC1} in a parameter space of $\Ra/\Ra_w$ and $\Ek$.  The transition to bulk rotating convection would occur in an infinite system via linear instability at $\Ra_c \approx \left (8.7-9.6 \Ek^{1/6}\right) \Ek^{-4/3}$ \cite{Chandrasekhar1961, Niiler1965}.  In the presence of sidewalls, however, the transition to a bulk convection state depends on $\Gamma$ and on the nonlinear state of the wall modes because of the non-zero base state \cite{Zhong1993} with $\Nu >1$.  

Whereas at modest $\Ek \gtrsim 10^{-5}$ the onset of bulk convection $\Ra_c/\Ra_w \lesssim 10$, for smaller $\Ek$ there is an expanding and more nonlinear range of wall modes.  For large enough $\Ra$ at fixed $\Ek$, buoyancy dominates over rotation, 
and the transition to this regime for $\Ek \lesssim 10^{-6}$ and $\Pran < 1$ is identified empirically as
$\Ra_t \sim \Ek^{-2}$ \cite{Zhang2021, Wedi2021}.
 The intermediate regime of bulk rotation-dominated convection is known as the geostrophic regime where convective Taylor columns, vortical plumes, and the condition of geostrophy are of great interest.  To understand experiments and DNS in realistic, confined convection cells, it is crucial to characterize the role of wall modes on the nonlinear evolution from no convection into the geostrophic regime and to connect the wall modes with the recently discovered boundary zonal flow (BZF) that exists in the turbulent geostrophic regime \cite{Zhang2020,Wit2020,Favier2020}.  That is our task here.

\begin{figure*}
\includegraphics[width=18cm]{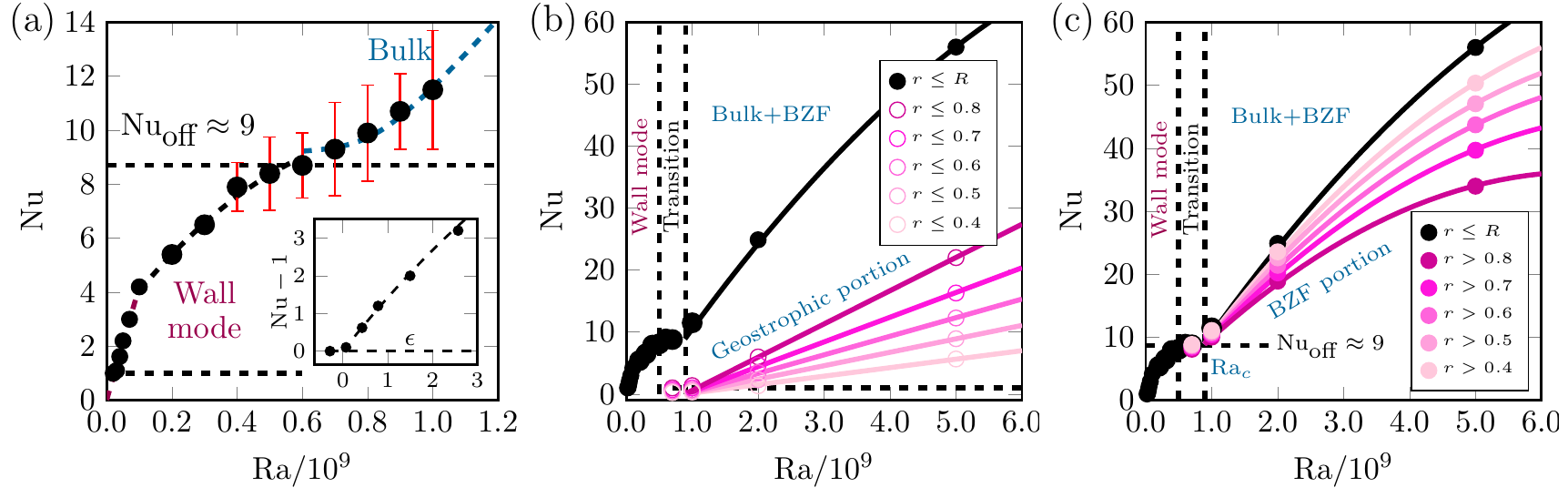}
\caption{$\Nu$ vs. $\Ra$ for $\Pran=0.8$, $\Ek=10^{-6}$ and $\Gamma=1/2$. 
(a) Region of pure wall modes $\Ra < 5 \times 10^8$ and  onset of bulk convection for $\Ra \gtrsim 9 \times 10^8$. Vertical bars are standard deviations of $\Nu$ fluctuations.   
$\Nu_\text{off}$ is the amount contributed by wall modes at bulk convection onset. 
(b),(c) Larger range of $\Ra$ with total $\Nu$ (solid, black) and contributions averaged over regions defined by (a) $r/R \leq r_0$ (bulk modes) and (b) $r/R > r_0$ (wall modes/BZF).}
\label{PIC3}
\end{figure*}
\begin{figure*}
\includegraphics[width=18cm]{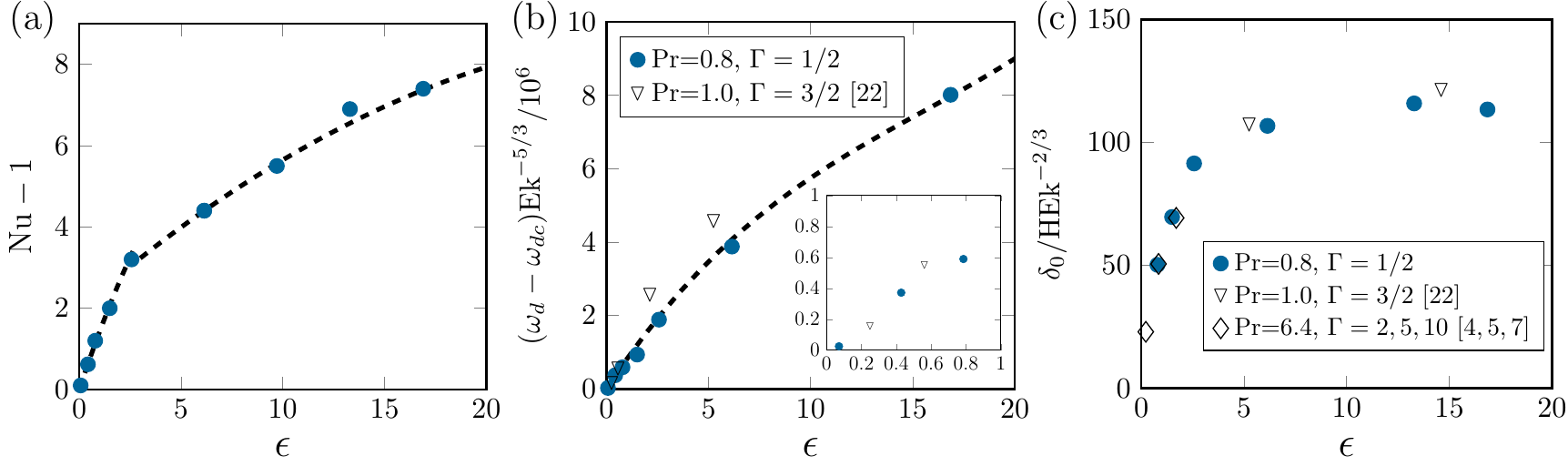}
\caption{
(a) Nusselt number, (b) $\left (\omega_d - \omega_{d_c}\right ) \Ek^{-5/3}$, and (c) $(\delta_0/H) \Ek^{-2/3}$ vs.\,$\epsilon$, for $\Pran$ = 0.8 and $\Ek = 10^{-6}$. Also plotted (open triangles) in (b) and (c) data from \cite{Favier2020}, and (open diamonds) in (c) data from \cite{Zhong1993, Ning1993, Liu1999} for wall-mode radial width from shadowgraph for different $\Gamma = 2, 5, 10$, $\Pran = 6.4$ with $\Ek = 2.3 \times 10^{-4}, 2.3 \times 10^{-4}, 1.8 \times 10^{-3}$ and $\epsilon = 0.84, 1.7, 0.22$, respectively.
}
\label{PIC4}
\end{figure*}
\begin{figure*}
\includegraphics[width=18cm]{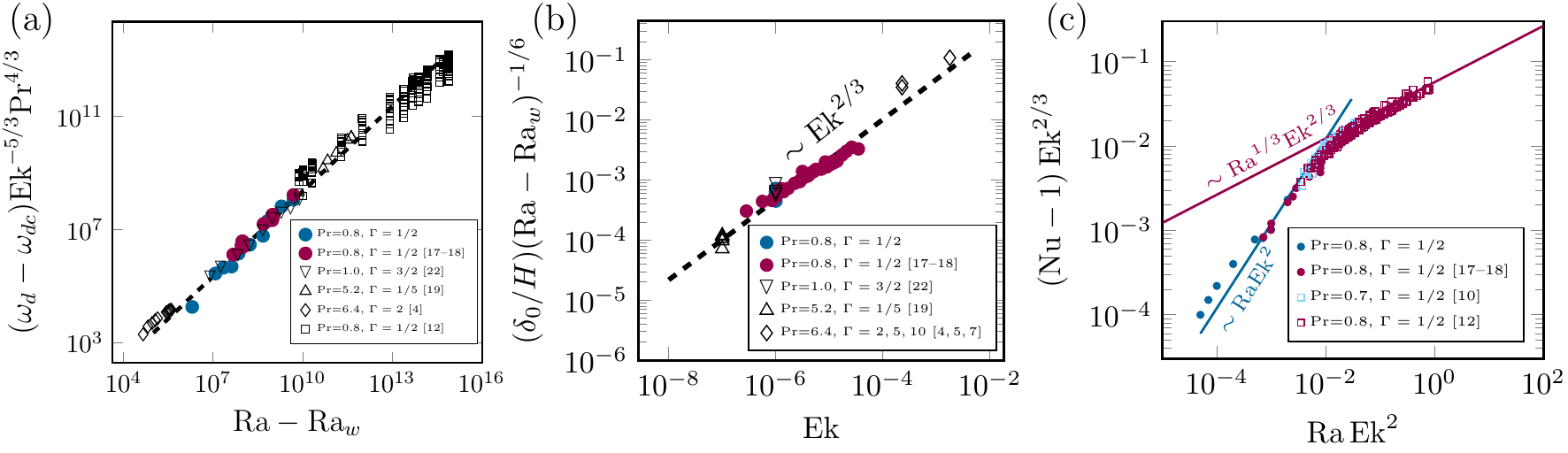}
\caption{
 Scaled (a) boundary mode drift frequency $\left (\omega_d - \omega_{d_c}\right) \Ek^{-5/3} \Pran^{4/3} \approx 0.022 \left (\Ra - \Ra_w \right)$,
(b) sidewall boundary length scale $(\delta_0/H) \left (\Ra - \Ra_w \right )^{-1/6} \approx 4.7 \Ek^{2/3}$, and (c) heat transport $\left (\Nu-1\right )\Ek^{2/3}$ vs. $\Ra \Ek^2$.
}
\label{PIC5}
\end{figure*}

A qualitative understanding of the evolution of the state of rotating convection can be gained by considering instantaneous mid-plane horizontal cross sections of temperature fields and associated streamlines and corresponding vertical temperature fields at the sidewall boundary ($r =0.98R$).  Fig.~\ref{PIC2} shows these fields for $\Ek = 10^{-6}$ and for several $\Ra$; also labeled is the reduced $\Ra$ defined as $\epsilon = \Ra/\Ra_w - 1$ where we take the experimental value $\Ra_w = 2.8 \times 10^7$.  Very close to onset ($\epsilon \approx 0.07$), the flow is organized as a mode-1 state with symmetric upwelling warmer (red) and downwelling cooler regions (blue), an overall anti-cyclonic rotation at the mid-plane, and a sinusoidal mean-temperature isotherm in the vertical field as shown in Fig.~\ref{PIC2}(a) (the retrograde direction of precession in the vertical profiles is to the left). Because of the confined geometry with $\Gamma = 1/2$, the wall mode thermal field is largest near the boundary but extends significantly into the cell interior; this has important implications for the heat transport crossover from wall modes to bulk modes described later.  

With increasing $\Ra$, the wall-mode state becomes more nonlinear but time-independent (in a frame co-rotating with the retrograde traveling wall mode) for $\Ra \lesssim 4 \times 10^8$. The state presented in Fig.~\ref{PIC2}(b) for 
$\Ra = 5 \times 10^8$ shows the more complex horizontal temperature field and flow circulation and the strongly nonlinear square-wave-like vertical profile with forward/backward (left/right) asymmetry; it is also weakly time dependent suggesting a wall-mode transition to an oscillatory state. For larger $\Ra$, Fig.~\ref{PIC2}(c), the streamlines are irregular, indicating unsteady flow and thermal inhomogeneity appears in the interior. One sees vertical striations arising from the influence of aperiodic time-dependent bulk modes interacting with the wall mode; a weak BZF has appeared.   

The wall mode state is characterized by four main properties that we consider here: the heat transport $\Nu$, the precession frequency $\omega$, the azimuthal mode number, and the radial distribution of convective amplitude (as measured by heat transport or azimuthal velocity $u_\phi$).  The azimuthal mode number is 1 because of small $\Gamma = 1/2$.  Previously we demonstrated that for the BZF $m=1$ for $\Gamma \leq 3/4$ and $m=2\Gamma$ for $\Gamma = 1$ or 2 \cite{Zhang2021}.  Our data show continuity from wall mode to BZF but multiple mode-number states are stable for larger $\Gamma$ and further study is necessary to elucidate this relationship. 

We first consider the heat transport and its contributions from the wall mode, from the bulk state, and from the BZF.  In Fig.~\ref{PIC3}(a), we show $\Nu$ versus $\Ra$ that covers the wall mode regime $3 \times 10^7 < \Ra < 5 \times 10^8$, a transition region $5 \times 10^8 < \Ra < 9 \times 10^9$, and the onset of strong bulk modes coexisting with remnant sidewall-localized modes, i.e., a BZF.  The inset shows linear growth (with quadratic corrections) of the wall mode heat transport near onset consistent with the expected scaling $\Nu -1 = a \epsilon + b \epsilon^2$. The fit gives $a=1.5$ and $b=-0.08$ where $\Ra_w =2.8 \times 10^7$ (compared to the theoretical value $3.2 \times 10^7$ for an insulating sidewall and a planar wall \cite{Herrmann1993, Zhang2009}).  As the wall modes become more nonlinear, $\Nu$ increases less rapidly and approaches an inflection point around $\Ra \approx 5 \times 10^8$ where a weak signature of time-dependent convection can be detected (vertical bars denote root-mean-square fluctuations).  At slightly higher $\Ra \approx 7 \times 10^8$, bulk modes begin to grow as demonstrated in instantaneous horizontal and vertical slices. At $\Ra \approx 10^9$, $\Nu$ increases rapidly as bulk convection and wall localized convection act together.  In Fig.~\ref{PIC3} (b,c), total $\Nu$ increases roughly linearly from an effective $\Nu_\text{off} \approx 9$ and $\Ra_c = 8.9 \times 10^8$ (compared to linear-stability prediction $\Ra_c \approx 7.8 \Ek^{-4/3} = 7.8 \times 10^8$ for $\Ek = 10^{-6}$ \cite{Niiler1965}).  Given the nonlinear base state created by the wall modes, the correspondence for the onset of bulk convection is good.  The nonlinear wall mode might round the transition to the bulk mode state, an idea  illustrated with the dashed blue lines in Fig.~\ref{PIC3}(a), consistent  with a fit to an imperfect bifurcation.  

To further explore the relative contributions of the wall localized states and the bulk state, we divide up the heat transport according to a radial separation $r_0$ (normalized by cell radius $R$).  
We denote the portion from 0 to $r_0$ as contributing to the bulk state whereas the remaining portion from $r_0$ to 1 is attributed to wall states.  In the language of \cite{Zhang2020}, the wall portion is the nonlinear BZF whereas the bulk convection is rotation dominated and in the geostrophic regime.  Although the quantitative split between different regions depends on the choice of $r_0$, the trends are unambiguous.  The $\Nu$ portion attributed to bulk modes grows linearly (or slightly super linearly) whereas the BZF contribution is sub-linear.  Thus, the bulk mode becomes  relatively more important with higher $\Ra$ although the BZF portion remains larger throughout the range studied here. Notice that an effective manner in which to describe the data for $\Ra>Ra_c$ is $\Nu-\Nu_\text{off} = a (\Ra/\Ra_c -1) + b (\Ra/\Ra_c -1)^2$ as indicated by the colored lines in Figs.~\ref{PIC3}~(b,c). We choose $\Nu_\text{off}$ as its value at a $\Ra$ where $\Nu$ increases rapidly.

We now present the $\Ra$ dependence of $\Nu$, $\omega_d$, and $\delta_0$ in the wall mode region as shown in Fig.~\ref{PIC4}.  To emphasize the variation with respect to the wall-mode onset, we use $\epsilon = \Ra/\Ra_w - 1$ as the abscissa and plot the quantities $\Nu - 1$, $\omega_d - \omega_{d_c}$, and $(\delta_0/H) \Ek^{-2/3}$ (the factor $\Ek^{-2/3}$ nicely collapses the data for $\Pran = 0.8$ and $\Pran=6.4$ \cite{Zhong1993}), respectively. Here we use the asymptotic linear stability result for a planar wall $\omega_{d_c} \approx  132\ \Ek/\Pran - 1464\ \Ek^{4/3}/\Pran$ \cite{Herrmann1993,Zhang2009}.  The trends show the nonlinear evolution of the wall mode states with increasing $\Ra$.

We now show trends in $\Nu$, $\omega_d$ and $\delta_0$ extending over many decades in $\Ra$ that spans the wall mode region and the geostrophic region coexisting with the BZF.  
In Figs.~\ref{PIC5}(a,b) we show scaled quantities 
$\left (\omega_d - \omega_{d_c}\right) \Ek^{-5/3} \Pran^{4/3} \approx 0.022 \left (\Ra - \Ra_w \right)$ 
(consistent with \cite{Zhang2021})
and $(\delta_0/H) \left (\Ra - \Ra_w \right )^{-1/6} \approx 4.7 \Ek^{2/3}$  
where we have combined data reported here and from \citep{Zhang2020,Zhang2021} with data from \cite{Zhong1993,Ning1993,Liu1999,Wit2020,Favier2020,Wedi2021}. 
The $\Ra$ and $\omega_d$ dependences are corrected for their finite values at the onset of wall modes using $\Ra-\Ra_w$ and $\omega_d-\omega_{d_c}$, respectively.  
The collapse over almost 10 decades in $\Ra-\Ra_w$ (Fig.~\ref{PIC5}a) showing 
$\omega_d-\omega_{d_c} \sim \Ra-\Ra_w$ 
and over 4 decades in $\Ek$ showing $\delta_0 \sim \Ek^{2/3}$, see Fig.\ \ref{PIC5}b, unambiguously establishes the connection between the wall modes and the BZF.

Finally, we consider the scaling of $\Nu$ where we plot $(\Nu-1) \Ek^{2/3}$ versus $\Ra \Ek^2$  in Fig.~\ref{PIC5}(c).  The data for $\Pran \approx 1$ collapse very nicely.  We choose this scaling arrangement to reveal the $\Nu-1 \sim \Ra^{1/3}$ independent of $\Ek$ for large $\Ra$, and $\Nu-1 \sim \Ra \,\Ek^{4/3}$ at smaller $\Ra$.  As demonstrated earlier in Fig.~\ref{PIC3}(b), a linear dependence of the geostrophic contribution to the total heat transport is obtained by subtracting the wall mode contribution $\Nu_\text{off}$ and taking the bulk reduced $\Ra$ as $\epsilon_b = \Ra/\Ra_c -1$, similar to analysis in \cite{Zhong1993,Ning1993}.  Other scalings can be confusing when ignoring the wall mode contribution which has mostly been overlooked in recent experiments on the geostrophic regime.  Further investigation of the scalings presented here, the nature of the nonlinear wall mode states, the imperfect crossover to bulk RRBC modes, and the role of $\Gamma$ on the properties reported here are important to fully characterize RRBC.

The authors acknowledge support from the Deutsche Forschungsgemeinschaft (DFG), SPP~1881 "Turbulent Superstructures" and grants Sh405/7 and Sh405/8, from the LDRD program at Los Alamos National Laboratory and by the Leibniz Supercomputing Centre (LRZ).


%

\end{document}